\documentclass[a4paper,11pt]{article}
\usepackage{pos}

\def\beqn{\begin{eqnarray}} \def\eeqn{\end{eqnarray}}
\def\beq{\begin{equation}} \def\eeq{\end{equation}}

\title{
\vspace*{-1.5cm}
\begin{minipage}{\textwidth}
{\normalfont\small 
\hspace{\fill} August 2024
}\\
\end{minipage}\\[60pt]
  Local renormalisation from Causal Loop-Tree Duality}

\ShortTitle{Local renormalisation from Causal Loop-Tree Duality}

\author*[a]{German F. R. Sborlini}
\author[a]{Jos\'e A. R\'ios-S\'anchez}

\affiliation[a]{Departamento de F\'isica Fundamental e IUFFyM, Universidad de Salamanca, 37008 Salamanca, Spain.}

\emailAdd{german.sborlini@usal.es}

\abstract{We report recent progress on the development of a local renormalisation formalism based on Causal Loop-Tree Duality (LTD). By performing an expansion around the UV-propagator in an Euclidean space, we manage to build counter-terms to cancel the non-integrable terms in the UV limit. This procedure is then combined with the so-called causal representation, and the UV expansion is performed at the level of on-shell energies. The resulting expressions are more compact, and they retain nice properties of the original causal representation. The proposed formalism is tested up to three-loops, with relevant families of topologies. In all the cases, we successfully cancel the UV divergences and achieve a smooth numerical implementation. These results constitute a first step towards a new renormalisation program in four space-time dimensions (by-passing DREG), perfectly suitable for fully numerical simulations.}

\FullConference{42nd International Conference on High Energy Physics (ICHEP2024)\\
18-24 July 2024\\
Prague, Czech Republic\\}

\begin{document}
\maketitle

\section{Introduction to Causal Loop-Tree Duality}
\label{sec:Introduction}
To overcome the challenges in high-energy physics, several new and efficient techniques have been developed in the last years. Some techniques, such as the Loop-Tree Duality (LTD), focus on the local cancellation of divergences to ease and speed-up the numerical implementations. Conceptually, LTD \emph{transforms virtual particles into real ones}: loop amplitudes are expressed as tree-level-like structures integrated over real-radiation phase-spaces. Since its introduction in 2008, LTD has witnessed major advances, incorporating very recently a reformulation that naturally preserves causality: the so-called Causal Loop-Tree Duality (cLTD) \cite{Verdugo:2020kzh,Capatti:2020ytd}. According to this approach, any $L$-loop scattering amplitude with $V$ vertices can be written as
\beq 
{\cal A}^{(L)} = \sum_{\sigma \in \Sigma} \int_{\vec{\ell}_1, \ldots, \vec{\ell}_1} (-1)^k\, \frac{{\cal N}_{\sigma}}{x_{L+k}} \, \prod_{i=1}^k \frac{1}{\lambda_{\sigma(i)}} \, + \, (\sigma \longleftrightarrow \bar{\sigma}) \, ,
\label{eq:CausalEQ}
\eeq
where $\lambda$ represents a physical or causal threshold of the amplitude and $\sigma$ indicates a combination of $k=V-1$ compatible causal thresholds. Since $\lambda$ contains same-sign combinations of positive on-shell energies, Eq. (\ref{eq:CausalEQ}) only contains \emph{physical} singularities.  

\section{Local renormalisation}
\label{sec:LocalRenormalization}
The main aim of the study conducted in Ref. \cite{Rios-Sanchez:2024xtv} is removing ultraviolet (UV) divergences from the Causal LTD representation through suitable Taylor expansions of the loop-momenta in the different UV limits. Inspired by the ideas explored in Ref. \cite{Becker:2010ng}, these expansions are carried out up to logarithmic order in the scaling parameters and naturally introduces the renormalisation scale. The expansion is applied to every possible combination of loop momenta, re-scaling them homogeneously, so that all simultaneous UV limits of internal lines are removed.

\begin{figure}[h!]
\centering
\includegraphics[scale=0.55]{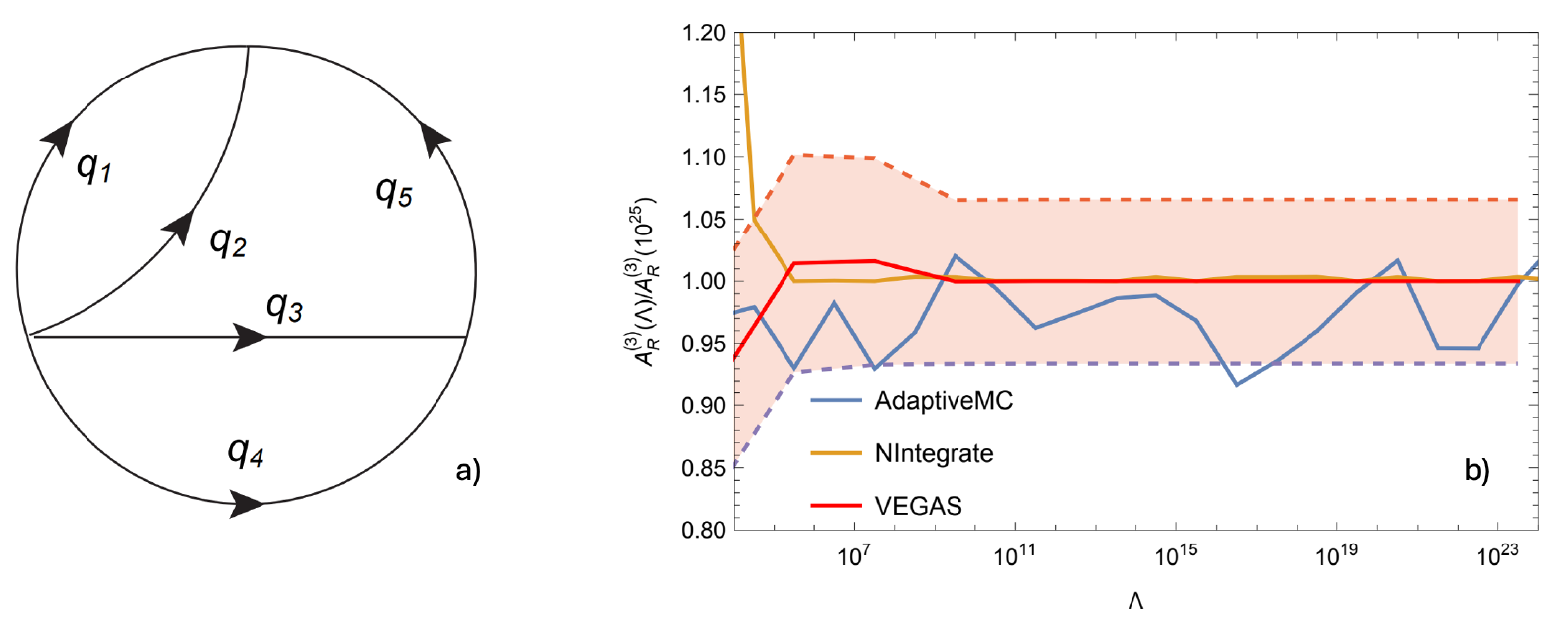}
\caption{Application example. (a) 3-loop NMLT diagram. (b) Analysis of the convergence of the locally renormalised amplitude, varying the cut-off scale $\Lambda$. Different numerical integrators are used in order to provide an error estimation.}
\label{fig:NMLTExample}
\end{figure}

To gain intuition on how this method works, let us consider the 3-loop next-to-maximal loop topology (NMLT), $A^{(3)}$, depicted in Fig. \ref{fig:NMLTExample} (left). The corresponding causal representation depends on the independent loop three-momenta $\{\vec{\ell}_1,\vec{\ell}_2,\vec{\ell}_3\}$. The first step consists in re-scaling and Taylor expanding each three-momenta, independently. This leads to the single UV-limit counter-terms, $A^{3}_{{\rm UV},i}$, and we define $(A^{(3)})'=A^{(3)}-\sum A^{3}_{{\rm UV},i}$. Then, we re-scale and Taylor expand two loop three-momenta simultaneously within $(A^{(3)})'$, obtaining the double UV-limit counter-terms, $A^{3}_{{\rm UV},ij}$. Finally, we define $ (A^{(3)})''=A^{(3)}-\sum A^{3}_{{\rm UV},ij}$ and we calculate the full local UV counter-term by simultaneously re-scaling all the loop three-momenta and Taylor expanding. In Fig. \ref{fig:NMLTExample} (right), we report the behaviour of the locally renormalised amplitude $A_{\rm R}^{3}$ as a function of the cut-off energy $\Lambda$. The calculation is implemented using different integration methods, in order to test the quality of the convergence. In the three scenarios considered, we find a very good and fast convergence: this is due to the perfect matching in the UV divergent behaviour of the original amplitude and our local counter-term. 

This iterative procedure allows to cancel the so-called overlapped singularities, originated in the simultaneous multiple UV limits. In this sense, we proof that our method is equivalent to the well-known BPHZ approach in the case of Feynman diagrams without disjoint sub-graphs, which justifies the aforementioned cancellation.

\section{Conclusions}
\label{sec:conclusions}
An integrand-level renormalisation method based on Causal Loop-Tree Duality (cLTD) was developed. It was tested on representative 2- and 3-loop amplitudes, finding in all cases a very smooth cancellation of UV divergences \cite{Rios-Sanchez:2024xtv}. In this article, we explained how to compute the local renormalisation counter-term for a next-to-maximal loop topology (NMLT) at three loops, and presented a convergence plot.

Besides leading to a very smooth convergence, our approach is compatible with BPHZ renormalisation which ensures the cancellation of overlapped singularities. This property makes our strategy particularly useful for removing UV singularities in the context of local methods, such as the recently proposed Casual Unitarity (CU) \cite{Ramirez-Uribe:2024rjg,LTD:2024yrb}. In the future, we expect to fully incorporate our local renormalisation approach within CU, and compute full cross-sections in four space-time dimensions at higher-perturbative orders.

\subsection*{Acknowledgments}
GS is partially supported by EU Horizon 2020 research and innovation programme STRONG-2020 project under Grant Agreement No. 824093 and H2020-MSCA-COFUND USAL4EXCELLENCE-PROOPI-391 project under Grant Agreement No 101034371.


\providecommand{\href}[2]{#2}\begingroup\raggedright\endgroup

\end{document}